# RobustAnalog: Fast Variation-Aware Analog Circuit Design Via Multi-task RL

Wei Shi[1], Hanrui Wang[2], Jiaqi Gu[1], Mingjie Liu[1], David Z. Pan[1], Song Han[2], Nan Sun[1]

[1]*University of Texas at Austin,* [2]*Massachusetts Institute of Technology*

## ABSTRACT

Analog/mixed-signal circuit design is one of the most complex and time-consuming stages in the whole chip design process. Due to various process, voltage, and temperature (PVT) variations from chip manufacturing, analog circuits inevitably suffer from performance degradation. Although there has been plenty of work on automating analog circuit design under the *typical* condition, limited research has been done on exploring robust designs under the *real and unpredictable* silicon variations. Automatic analog design against variations requires prohibitive computation and time costs. To address the challenge, we present RobustAnalog, a robust circuit design framework that involves the variation information in the optimization process. Specifically, circuit optimizations under different variations are considered as a set of tasks. Similarities among tasks are leveraged and competitions are alleviated to realize a sample-efficient multi-task training. Moreover, RobustAnalog prunes the task space according to the current performance in each iteration, leading to a further simulation cost reduction. In this way, RobustAnalog can rapidly produce a set of circuit parameters that satisfies diverse constraints (*e.g.* gain, bandwidth, noise...) across variations. We compare RobustAnalog with Bayesian optimization, Evolutionary algorithm, and Deep Deterministic Policy Gradient (DDPG) and demonstrate that RobustAnalog can significantly reduce required the optimization time by **14**×-**30**×. Therefore, our study provides a feasible method to handle various real silicon conditions. The code for RobustAnalog is available in this link.

## 1 INTRODUCTION

Analog circuit design is a paramount but extremely challenging task. It requires a huge amount of human efforts and lacks effective automations. Due to numerous chip manufacturing variations, analog circuits suffer from non-trivial performance degradation. Addressing such variation issues is considerably challenging. Large manufacturing variations make the circuit performance *unpredictable*.

As the chip fabrication technology advances, variation issues become even worse, leading to a larger chip failure rate. If such severe variation issues are not carefully handled, significant economic losses up to the billions of dollars will occur [17]. Hence, an effective variation-aware circuit design methodology is in high demand.

Traditional solutions to address such circuit variation issues primarily rely on laborious human expert involvement. Experts manually design the circuit based on *their expertise* and the feedback from *a large number of circuit simulations* and iterate the process until it passes all variation tests. However, the *burdensome analysis and slow simulations* make the manual design process considerably time-consuming.

Existing automated methods cannot address variation issues effectively. The black-box optimization algorithms [5, 15] and learning-based automation techniques [14, 21, 26, 27] are used to design circuits. However, they merely focus on the optimization under *the typical condition without variations*. Another kind of work involving variations focuses on improving the yield [13]. None of them can rapidly and systematically produce a robust design achieving real circuit performance targets. The variation-aware optimization is challenging in two aspects. First, the *simulation cost is prohibitively expensive* in order to get accurate variation effects under many test cases. Second, different variation conditions might conflict with each other which *significantly complicates the circuit optimization problem*. It will cost the solver much more time to find a feasible solution that meets all performance constraints.

To address the above challenges, in this work, we present RobustAnalog, an efficient variation-aware optimization framework for automatic analog circuit design. RobustAnalog largely reduces the simulation cost to design a robust analog circuit against variations. Here the variation-aware optimization is formulated as a multi-task reinforcement learning (RL) problem, where design for each variation condition is considered as one task. RobustAnalog includes two stages. At the first stage, we select a representative subset of tasks as the training set. Specifically, we group the tasks using clustering algorithm and choose one task per group to form the training task set based on their relative performance to the target performance. At the second stage, we leverage multi-task deep deterministic policy gradient (DDPG) [12] to train our RL agent with the selected tasks. During training, the critic model learns to predict values of state-action pair from each task and guides the actor to generate a better policy. To alleviate conflicting multi-task gradients, we apply PCGrad [29] to optimize actor and critic models. The code for RobustAnalog is available in this link.

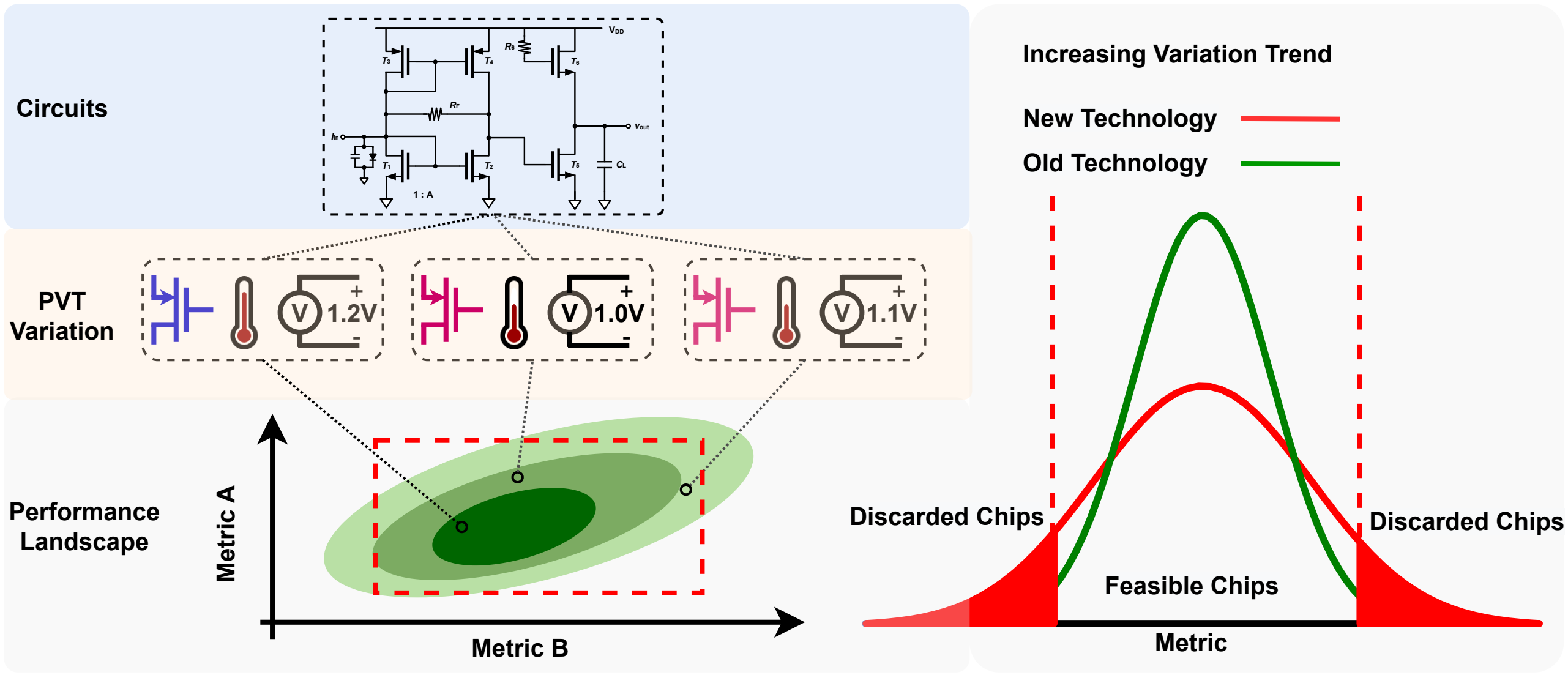

**Figure 1: *Left:* Performances under variations form a distribution. *Right:* New technologies have larger process variations, vulnerability to environmental variations, hence higher discarded rate.**

RobustAnalog tackles the problem of variation-aware automatic analog circuit sizing from several different perspectives. The core contributions of this work are as follows,

- We propose an automated optimization framework for variation-aware analog circuit design via multi-task reinforcement learning and adaptive task space pruning.
- An efficient training with variations is achieved by multi-task RL. The different PVT corners are formulated as multiple tasks. Sampling efficiency is largely improved by leveraging the correlations among similar tasks and mitigating the competition among conflicting tasks.
- An effective task pruning technique reduces the number of training tasks. The agent is trained with a series of small subset of tasks. Inheriting the trained weights from last cycle, our agent can be improved incrementally and achieve full tasks eventually. The number of queries into the full task set is minimized, leading to a significant simulation cost reduction.
- Extensive experimental results demonstrate that, on real-world circuit design benchmarks, our method outperforms Evolutionary strategy (ES), Bayesian optimization (BO), and DDPG methods with **14×-30×** simulation cost reduction.

## 2 PRELIMINARIES

**PVT Variation and Corners** – The major part of variations is *PVT variation*. PVT variation usually refers to a combination of global process variation (P), power supply (V), and temperature (T) variations. Process variations happen during chip manufacturing, resulting in different transistor characteristics. There are five transistor models to cover the process variation {TT, SS, FF, SF, FS}. To avoid circuit failures due to uncontrollable PVT variations, we model all these variations by a set of *PVT corners*. A PVT corner is a combination of process, voltage, and temperature values. For example, a fast-process, high-voltage, and low temperature corner is {Process = FF, $V_{dd}$ = 1.3V, T = 15°C }. A robust circuit should maintain desired performances in all of the pre-set PVT corners.

**Automatic Analog Sizing** – Automatic analog sizing techniques are attracting more and more research interests these years. The optimization methods, including Bayesian Optimization [15, 23], Genetic Algorithms [5] formulate the circuit design as a black-box problem. They show the differences in the sample efficiency and optimality. However, the critical issue is that they have to optimize the circuit from scratch every time when encountering a new design condition. The lack of tranferalibity across different conditions prevents them from addressing the variation issue at an affordable cost. Recently, learning-based methods have been extensively applied to circuit sizing problems. Deep neural networks (DNN) [26, 30] can approximate the complex relation between circuit parameters and performances. Deep RL methods show the potential to achieve higher circuit performances given enough explorations [21, 26, 27]. Moreover, RL enables transfer- learning across different design conditions, including different technologies and pre/post-layout design stage. However, current methods cannot reach the design goal under different conditions simultaneously. In this paper, our work effectively addresses the variations of real-circuits altogether.

**Multi-Task RL** – Deep reinforcement learning (DRL) is an emerging subfield of RL that can scale RL algorithms to complex and rich environments. Multi-task RL focuses on enabling the single agent to solve multiple related problems, either simultaneously or sequentially [25]. Learning multiple related tasks together should facilitate the learning of each individual task [1, 4]. However, it has also been found that training on multiple tasks can negatively affect performance on each task. Different kinds of techniques are proposed to solve this issue including new architectures [6, 9], auxiliary tasks [11], and new optimization schemes [10, 29]. Besides, choosing which task or tasks to train on at each time step is also important. The task scheduling [22] is also discussed. The idea behind it is to assign task scheduling probabilities based on relative performance to a target level. Optimized training task selections can significantly improve model performance [2]. We explored both the optimal task selection and multi-task training. They are integrated together into the framework to boost the sampling efficiency.

## 3 PROPOSED PVT VARIATION-AWARE CIRCUIT SIZING

### 3.1 Problem Definition

Given a fixed circuit topology, we search for a circuit sizing vector whose performance can satisfy the constraints (design targets) across all variations. Then the problem can be formulated as a constraint satisfaction problem under different conditions.

$$\begin{aligned} \text{minimize} \quad & 0 \\ \text{subject to} \quad & F_i(X|T_j) < C_i, \quad j = 1, \ldots, k \end{aligned} \tag{1}$$

where

$$\begin{aligned} X &\in D \\ X &= X_1, X_2, ..., X_n \\ D &= D_1, D_2, ..., D_n \\ C &= C_1, C_2, ..., C_m \\ T &= T_1, T_2, ..., T_k \end{aligned}$$

$X$, the sizing vector, is an $n$-dimensional variable which corresponding to $n$ circuit sizing parameters. $D$ is the domain for $X$. For example, $D_1$ is [0, 1] which means the design space of $X_1$ is [0, 1]. $T$ is the set of $k$ pre-defined PVT corners to cover possible variations in the real world. $C$ is the constraint set for all circuit metrics. Because we have $m$ metrics, the number of constraints is also $m$. $F_i(X|T_j)$ is the $i^{th}$ performance metric of circuit under the $j^{th}$ corner. $F_i$ is a non-linear mapping between $X$ and the $i^{th}$ metric in the performance. $X$ is the input, and $T$ is the parameter. We rely on the circuit simulator to provide this mapping. Therefore, our goal is to find an $X$ that can satisfy any constraints in $C$ under any corner task in $T$. It is worth noting that choosing which tasks to optimize is also non-trivial. Spending simulations on each task is wasteful and provides minimal additional information since the correlation among tasks is ignored. A more interesting way is to conduct the task selection and multi-task training jointly.

**Algorithm 1:** Multi-task RL in RobustAnalog

Given critic network $Q(S, A, Z_i \mid \theta^Q)$ and actor network $\mu(S \mid \theta^\mu)$ with critic weights $\theta^Q$ and actor weights $\theta^\mu$ ;
Given replay buffers $\{P_i\}$ ;
**for** *episode = 1, M* **do**
  Initialize random process $\mathcal{N}$; Reset all environments $S$;
  **if** *episode* $\leq W$ **then**
    Warm-up: randomly sample an action $\widetilde{A}$;
  **end**
  **else**
    Select action $\widetilde{A} = \mu(S \mid \theta^\mu) + \mathcal{N}$ according to the current policy and exploration noise;
  **end**
  Denormalize and refine $\widetilde{A}$ with design constrains to get $A$;
  Simulate the $A$ for each task to get rewards $\{R_i\}$ ;
  Store each transition $(S, A, R, Z_i)$ in $P_i$;
  **if** *episode* $> W$ **then**
    Sample a stratified batch of $(\widehat{S}, \widehat{A}, \widehat{R}, \widehat{Z}_i)$ from $\{P_i\}$ (batch size = $N_s$);
    Update the critic by minimizing K losses with PCGrad:
    $L_i = \frac{1}{N_s} \sum_{k=1}^{N_s} (\widehat{R}_k - B - Q(\widehat{S}_k, \widehat{A}_k, \widehat{Z}_i \mid \theta^Q))^2$;
    Update the actor using the K gradients modified by PCGrad:
    $\nabla_{\theta^\mu} J_i = \frac{1}{N_s} \sum_{k=1}^{N_s} \nabla_a Q(S, A, Z_i|\theta^Q)|_{\widehat{S}_k, \mu(\widehat{S}_k)} \nabla_{\theta^\mu} \mu(S|\theta^\mu)|_{\widehat{S}_k}$
    $\nabla_{\theta^\mu} J_i = PCGrad(\nabla_{\theta^\mu} J_i)$
  **end**
**end**

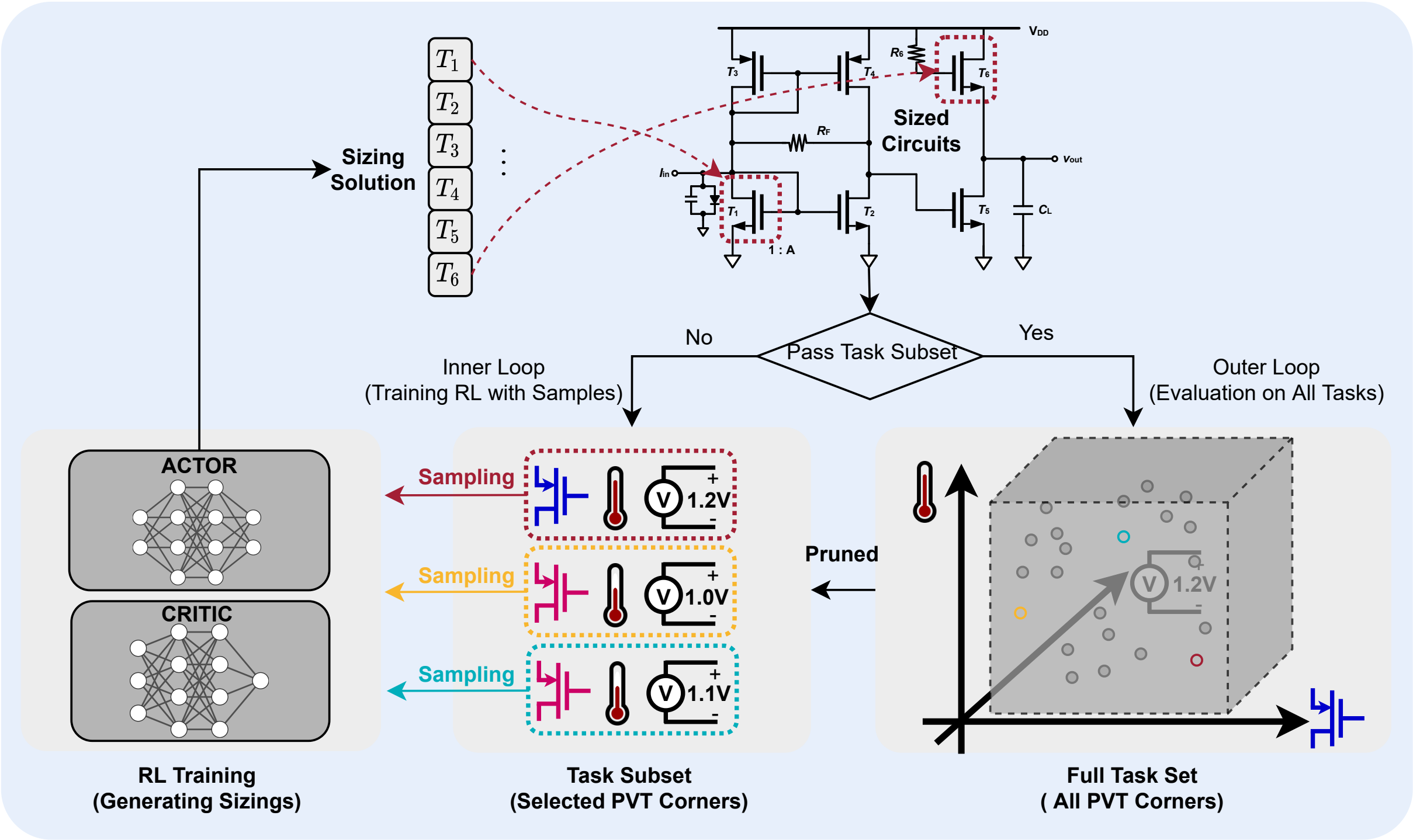

**Figure 2: RobustAnalog Overview. (1) A pruned task subset is generated from the full task set (2) Multi-task RL agent is trained on task subset (3) Training continues until the produced sizing can achieve training tasks. Then the sizing is evaluated on the full set. If it passes all the tasks, RobustAnalog returns the result.**

## 3.2 Framework Overview

An overview of the proposed framework is shown in Figure 2. We consider satisfying constraints under one PVT corner as a single task. In each iteration, (1) RobustAnalog selects a new task subset from all PVT corner tasks. For the first iteration, a pre-defined nominal corner will be selected as the first task; (2) The RL agent generates actions and passes them to each environment in the task subset; (3) The environment denormalizes actions ([-1, 1] range) to actual circuit sizings and refines them. The sizings will be truncated according to minimum precision, lower and upper bounds of the technology node if necessary; (4)Simulate the circuit (5) Agent gets the rewards from corner-specific environments. Optimizations are performed on the actor and critic networks with PCGrad technique. (6) If all tasks in the subset are passed during the agent evaluation, the sizing solution will be tested on the full task set. If it passes all tasks, the loop terminates. Otherwise back to (1). In the meantime, actor-critic model weights and replay buffers are saved for the agent to inherit in the next iteration.

## 3.3 Multi-Task RL training

Multi-task RL is a training paradigm in which the agents are trained with samples from multiple tasks simultaneously. Shared representations are learnt from a collection of related tasks. These shared representations increase sample efficiency and can potentially yield a faster learning speed for related tasks. In our setting, we create a multi-task agent whose critic can predict the value of task-conditioned action-state pairs. Since the target of the actor is to look for a sizing that passes all tasks, the actor model is set to be task agnostic. Another benefit from shared representations is its ability to generalize to unseen corner tasks, which is useful in Monte Carlo corner tests. There are more discussions in Section 4.

**State.** The PVT information is embedded in our states, $s = (p, v, t)$, where $p$ is the one-hot representation of component type, $v$ is the normalized voltage value and $t$ is the normalized temperature value.

**Reward.** Our reward is formulated as:

$$R = \begin{cases} r, & r < -0.02 \\ 0.2, & r \geq -0.02 \end{cases} \tag{2}$$

$$r = \sum_{i=1}^{M} \min\{\frac{m_i - m_i^*}{m_i + m_i^*}, 0\} \tag{3}$$

where $m_i$ is the current simulated $i^{th}$ performance metric and $m_i^*$ is the corresponding constraint. The reward is a measure of the relative distance between the current performance metrics and the corresponding design targets. Once the requirements are met, the reward value is fixed at 0.2. The value of 0.2 is fixed for any circuit without any tuning. This reward formulation is motivated by the design goal in the real world. Designers tend not to over-optimize the circuits. It is more important that designers can fulfill the requirements in a short period of time.

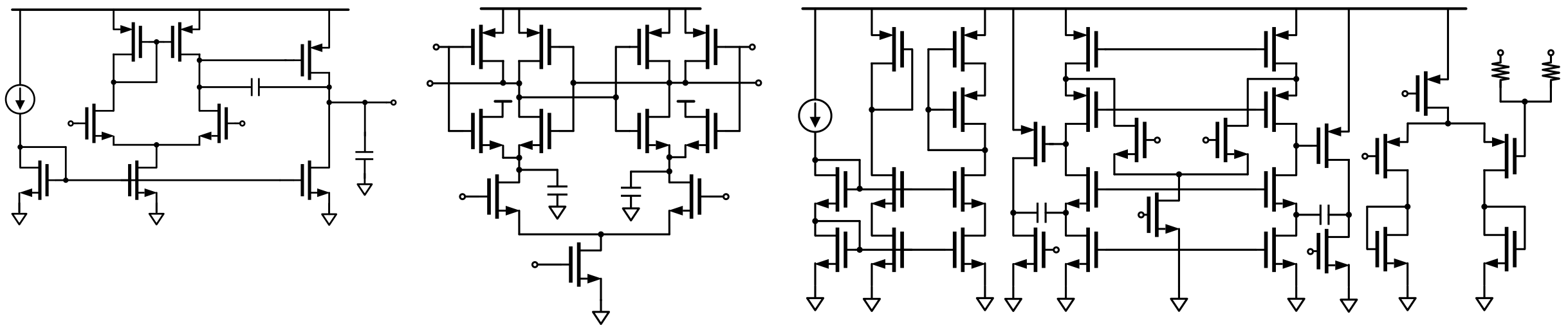

**Figure 3: Three analog/mixed-signal benchmarks.**

**Action.** The action vector is a set of values corresponding to the sizing parameters for each circuit. They include transistor sizes (width, length) and capacitor values. The details of settings for each benchmark are illustrated in section 4.

**Training.** The environment includes the circuit, simulator, and PVT information. Each time we query the environment, it simulates the circuit and returns the performance with PVT information. After agent-environment interactions, samples $(s, a, r, z_i)$ will be stored in the replay buffers , where $s$ is the state, $a$ is the action, $r$ is the reward, and $z_i$ is the corner task ID. The critic neural network takes $(s, a, z_i)$ as a input and predicts the corresponding value for the current corner task. Relying on the insight that performance under different corners are related, most of critic neural network parameters are shared across tasks except a few in the input layer. The task ID is removed from the inputs of the actor neural network. The training process is modified from DDPG [12]. Details are illustrated in Algorithm 1. $M$ is the max optimization episodes and $W$ is the warm-up episodes. $N$ is the truncated norm noise. $N_s$ is the training batch size. The key difference from the single-task setting is that we sample a stratified batch from buffers every time and generate task-specific losses. Also, samples from different tasks are stored in separated tasks. For the optimization strategy, we use PCGrad [29] to address conflicting gradients from different tasks.

## 3.4 Task Space Pruning

Although multi-task training has improved the efficiency of optimization on different corner tasks, we can still reduce the number of simulations further by selecting a small-sized training task set. Since our final goal is to pass all the corner tasks, we must be able to iteratively improve our optimization results with a series of training task sets, as shown in Fig. 2. Therefore, we choose to incrementally train our NN-based RL agent since it has the transferability and capability of inheriting the trained weights from last cycle[21, 26]. Such transferability makes it compatible with the following task space pruning technique, which is a major advantage of multi-task RL methods over other optimization methods like Bayesian optimization and Evolutionary strategy.

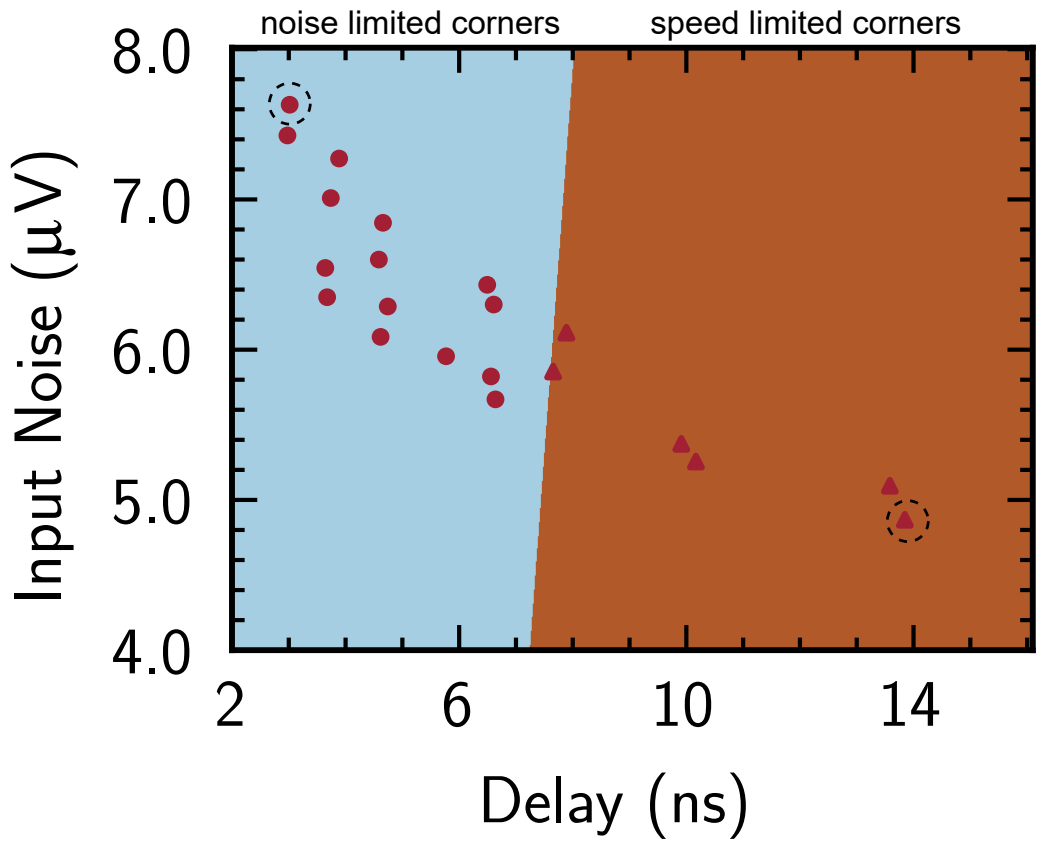

**Figure 4: Visualization of corner task clustering and selection for strongARM Latch. k-means decision boundary is shown, dividing corners into two kinds: noise-limited corners (blue) and speed-limited corners (brown). The corner with the worst performance in each cluster is chosen as one of the training tasks.**

Choosing a small batch from a large number of tasks is non-trivial. One straightforward way to form a training task set is sampling tasks randomly from the full set[7, 20]. For manual designs, human designers tend to guess the worst-case corners and design against those corners. Inspired by human design methodology, the work of [28] defines the lowest-reward as the worst cases and optimize on them correspondingly. The value of reward, scalarized from a multi-dimensional performance metric vector, lacks the information to differentiate different low-performance corners. Low rewards may result from different constraint violations. Therefore, the lowest reward doesn't mean one corner's performance is dominated by others'.

To address the problem mentioned above, we first cluster the corners based on their multi-dimensional metric vector and rank the corners in the same cluster by their rewards. Corners in the same cluster have the similar performance pattern. Therefore, the value of rewards in the same cluster can better reflect the "goodness" of one corner's performance. Since corner performance patterns are unknown, we perform clustering by using one of unsupervised learning techniques, k-means [16], in the performance space. After including all the performance patterns, we can train our agent with a small subset of all corners with confidence that all corners will still work. We also apply the pruning technique to larger random corner set, which is detailed in section 4.

We take strongARM benchmark as an example. From last cycle, we have an optimized sizing based on the last batch of training corners. If such optimized sizing doesn't pass the full corner test set, we need to select a new batch of training corners. The proposed task space pruning contains three steps: (1) We simulate the optimized on all corner tests to get the corresponding performance distribution. (2) We divide the corners into different clusters by using the performance metrics as input features. It is shown in Fig. 4 that corners of strongARM belong to two clusters. One cluster is noise limited and the other is speed limited. (3) We select the corner with the lowest reward in each cluster as one of the training tasks for the next iteration. Cluster-specific worst corner sets a lower bound of performances within the same group. With this pruning technique, the task space for multi-task RL training in each iteration is pruned to be a significantly smaller scale while still being a good representation of the full task space. If there is no sizing given at first, a pre-defined nominal corner will be chosen as the first corner to train on. An interesting finding from the empirical study results is that the easier corner tasks help to accelerate the learning of other hard tasks. Therefore, we always add a nominal corner as an auxiliary task in the training task sets at all time steps. If all the corners are passed, the loop terminates.

# 4 EXPERIMENTS

## 4.1 Analog/Mixed-signal Circuits

We experiment with three real-world analog/mixed-signal circuits. They are two-stage operational transimpedance amplifier (Two-stage OTA), folded-cascode operational transimpedance amplifier (Folded-Cascode OTA) and strongARM Latch. They are chosen for three reasons. First, they are the most important and common-used blocks in various systems. Engineers usually spend the longest time optimizing the performance and robustness of these circuits. Second, they include two representative kinds of analog circuits which are the static and dynamic circuits. The two kinds are dictated by different physic and engineering rules. The third reason is that they have different levels of variations. Two-stage OTA is with 45nm, and the other two are with older 180nm technology. 45nm has a larger variation. Therefore, we can study the impacts of different variation magnitudes. Each circuit is a composition of a number of transistors and capacitors. Each transistor has two parameters, the gate width and length ($w$, $l$). Capacitors have one parameter (c), the capacitance value. The initial design spaces of these devices are given by human designers. To minimize the efforts of designers, our design space are set to be very large. They have $10^{14}$, $10^{27}$, and $6.4 \times 10^{64}$ possible values correspondingly.

The circuits are simulated on SPICE-based simulators [18]. Two-stage OTA is on Ngspice and BSIM 45nm predictive technology [3]. Folded-Cascode OTA and strongARM Latch are on Cadence spectre and TSMC 180nm technology, a commercial simulator tool.

**Two-stage OTA.**. The topology is shown in Figure 3. It has 7 parameters including 6 transistor widths (w) and 1 capacitor value (c). The range of w is $[0.5, 50]*1\mu m$ and $[0.1, 10]*1pF$ for c. The total design space is $10^{14}$ possible values. The performance metrics are current(i), unity gain-bandwidth (ugb), phase margin (phm). The corresponding constraints (C) and the PVT corner tests (T) are showed below. There are 30 corners ($5 \times 3 \times 2$).

$$T = \{TT,\ SS,\ FF,\ FS,\ SF\} \times \{1.0V,\ 1.1V,\ 1.2V\ \} \times \{0^{\circ}C,\ 100^{\circ}C\ \}$$
$$C = \{i \leq 5mA,\ ugb \geq 15MHz,\ phm \geq 60^{\circ}\}$$

**Folded-Cascode OTA.**. The topology is shown in Figure 3. It has 20 parameters, including 7 transistor widths (w), 7 lengths (l), 2 capcitor values (c) and 4 transistor ratios (n). The range of w is $[0.24, 150]*1\mu m$, $[0.18, 2]*1\mu m$ for l, $[0.1, 2]*1pF$, $[0.1, 10]*pF$ for different c. The total design space is $6.4 \times 10^{64}$ possible values. The performance metrics are power(p), unity gain (g), phase margin (phm), common-mode rejection ratio (CMRR), power supply rejection ratio (PSRR), noise (n), unity-gain-bandwidth (ugb). The corresponding constraints (C) and the PVT corner tests (T) are showed below. There are 20 corners ($5 \times 2 \times 2$).

$$T = \{TT,\ SS,\ FF,\ FS,\ SF\} \times \{1.6V,\ 1.8V\ \} \times \{0^{\circ}C,\ 100^{\circ}C\ \}$$
$$C = \{p \leq 1mW,\ ugb \geq 30MHz,\ phm \geq 60^{\circ},\ n \leq 30mV,\ g \geq 60dB,\ CMRR \geq 80dB,\ PSRR \geq 80dB\ \}$$

**strongARM Latch.**. The topology is shown in Figure 3. It has 7 parameters, including 6 transistor widths (w), 1 capcitor values (c). The range of w is [0.22, 50]$*1\mu m$, [0.15, 4.5]$*1pF$ for c. The total design space is $10^{27}$ possible values. The performance metrics are power(p), set delay (sd), reset delay (rd), set voltage (sv), reset voltage (rv), noise (n). The corresponding constraints (C) and the PVT corner tests (T) are showed below. There are 20 corners ($5 \times 2 \times 2$).

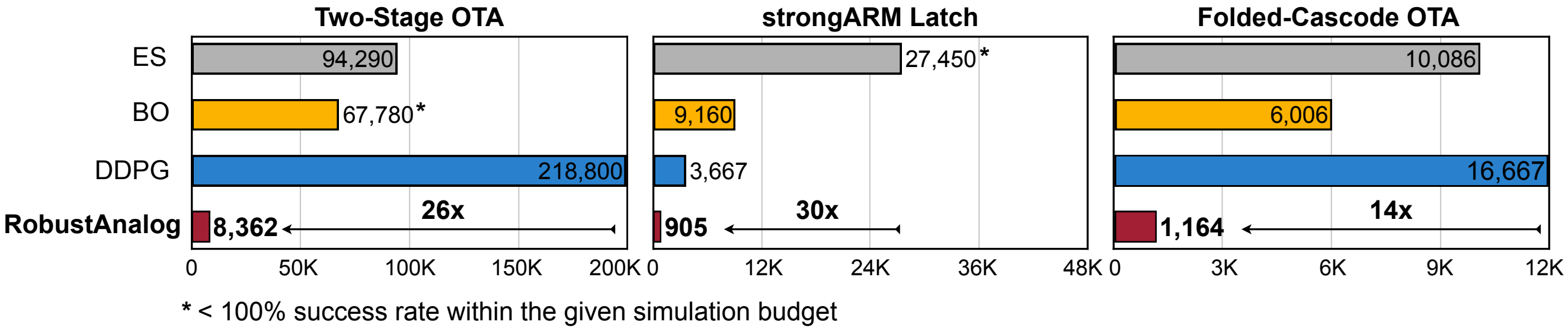

**Figure 5: Simulation times for each method to take to first hit reward=0.2**

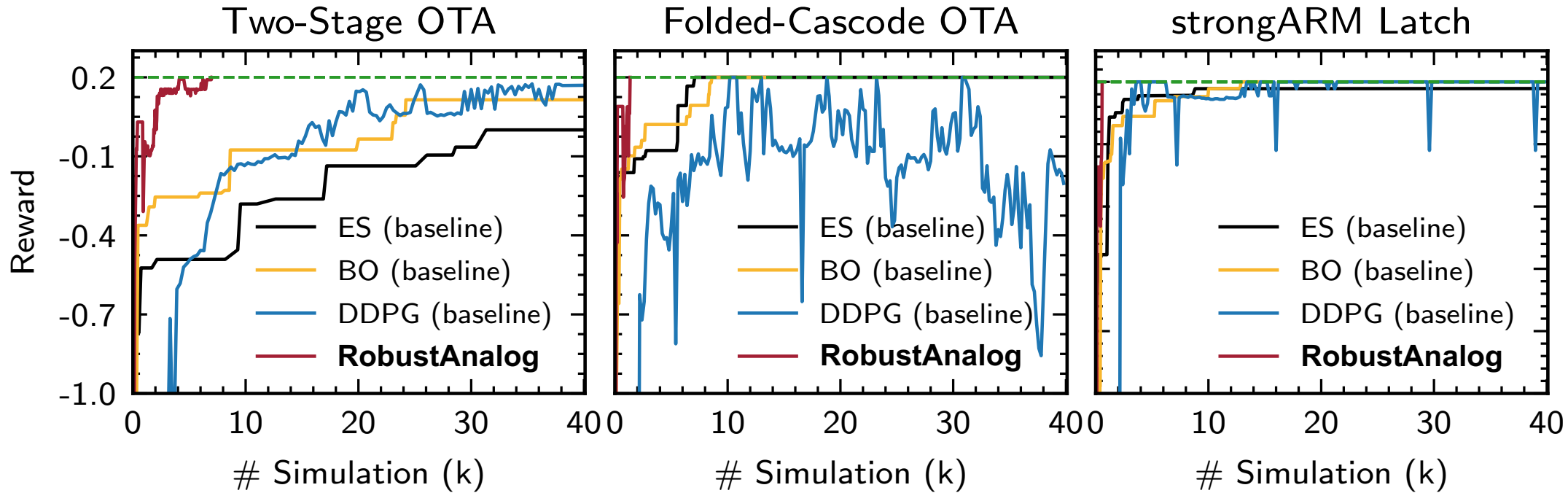

**Figure 6: Compare learning curves (average reward vs. # simulation) among baselines and our proposed RobustAnalog. Reward=0.2 indicates all tasks are passed. RobustAnalog hits the reward of 0.2 significantly faster than the baseline methods on all benchmarks.**

$$T = \{TT,\ SS,\ FF,\ FS,\ SF\} \times \{1.1V,\ 1.2V\ \} \times \{0^{\circ}C,\ 100^{\circ}C\ \}$$
$$C = \{p \leq 4.5uW,\ n \leq 50uV,\ sd \leq 14ns,\ rd \leq 9.1ns,$$
$$sv \geq vdd - 0.05V,\ rv \leq 0.05V\ \}$$

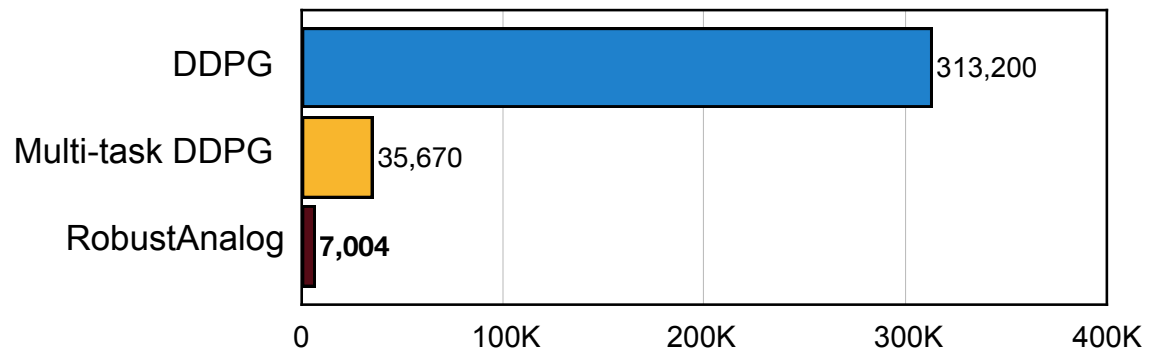

**Figure 7: Ablation of applying multi-task and task space pruning. Using two together brings the least simulation cost.**

## 4.2 Training Settings

To demonstrate the effectiveness of the proposed RobustAnalog, we apply RobustAnalog to the above three circuits and record the simulation time it took to pass all the corner tests. We compare the results of RobustAnalog with Bayesian Optimization (BO) [23], Evolutionary Strategy (ES) [8], and single-task RL algorithm (DDPG). For the three baselines, the variation-aware circuit optimization is considered as a single task. The average reward of all corner tasks is used to indicate the goodness of the current sizing. BO, ES, and DDPG improve the average reward until it reaches 0.2. In ES, DDPG, and RobustAnalog, the circuit simulation time accounts for over 95% of the total time. The computation time of BO becomes comparable with simulation time after many iterations. We compare these methods in terms of the simulation time. For RL training, we use a training batch size of 64, replay buffer size of 1000, and exploration noise standard deviation of 0.2. Actor and critic are all 4-layer multilayer perceptions (MLPs). For RL methods, we evaluate the agent every 10 training steps. All the experiments are conducted on a 6 core CPU. RL methods are implemented with PyTorch [19, 24]

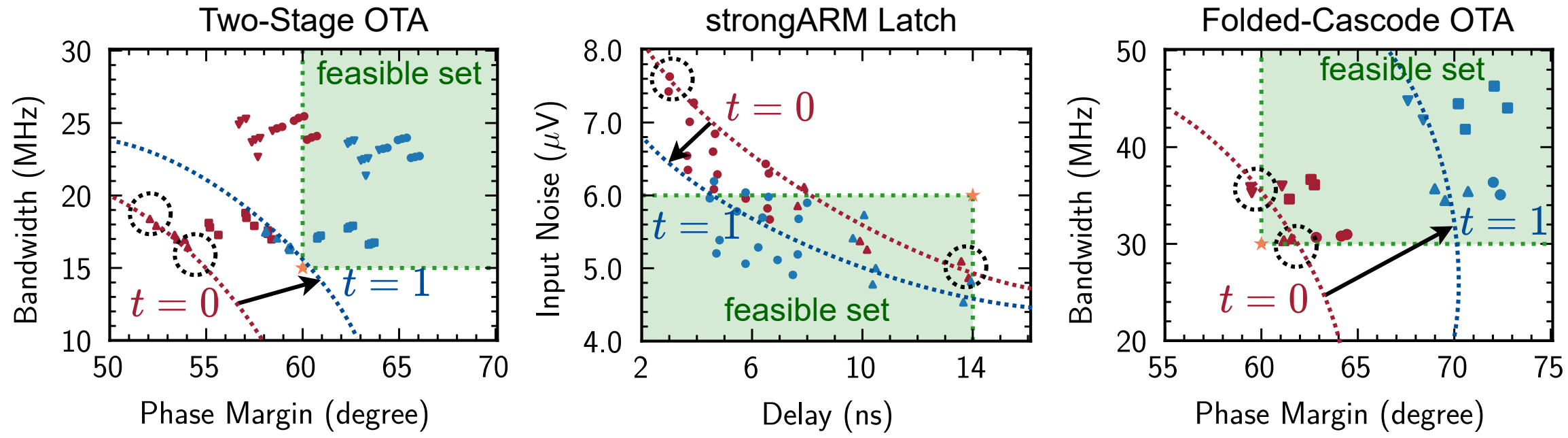

**Figure 8: Performance distributions of two intermediate sizings during the RobustAnalog optimization. Red and blue markers are performances on different corners at time $t_0$ and $t_1$. Selected training corners are indicated by black circles.**

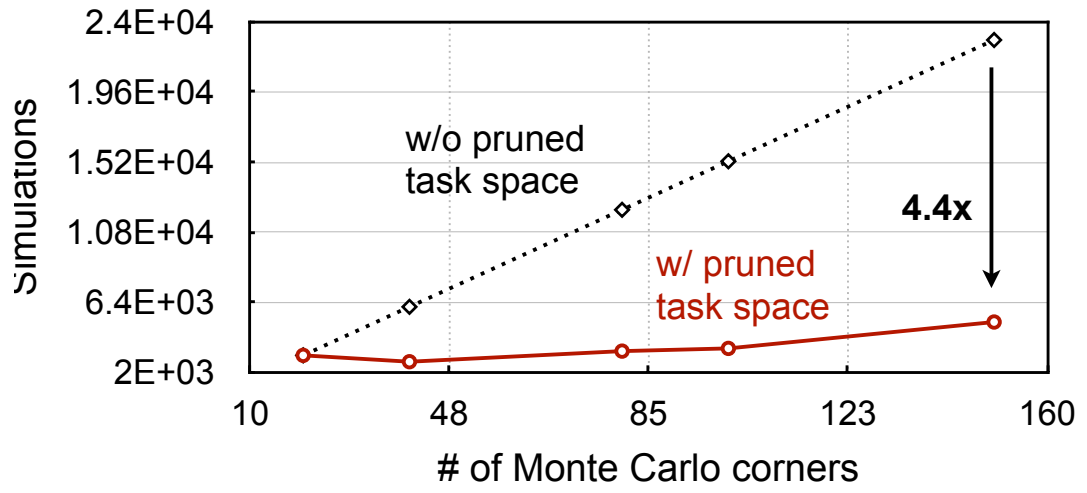

**Figure 9: Required simulation steps with more corners**

## 4.3 Evaluation of the Circuit Optimization

In all three circuit benchmarks, RobustAnalog achieved the smallest simulation cost to accomplish all the corner tasks. In each benchmark, it passed all the corners in the runs of different random seeds hence a 100% success rate. The comparison of simulation costs are shown in Figure 5. RobustAnalog consistently outperforms the baseline methods including ES, BO, and single-task DDPG. The simulation cost reductions are huge, 26x in Two-Stage OTA, 30x in strongARM Latch, and 14x in Folded-Cascode OTA. Note that BO becomes slow after having many samples. We ran BO for the same time with other methods for fair comparisons. We have several findings from the experiment results. First, all methods spend more simulations on optimizing the Two-Stage OTA which has larger variations with the 45nm technology. Second, compared to the ES and BO, single-task DDPG performs better in strongARM Latch while worse in the Two-Stage and Folded-Cascode OTAs. This is possibly because strongARM Latch is a dynamic circuit that is different from the static OTAs. To conclude, RobustAnalog shows a significant efficiency improvement in the different levels of variations and circuit benchmarks with distinct natures. The learning curves are shown in Figure 6.

## 4.4 Analysis

**Multi-Task and Task Space Pruning..** We conduct an ablation study on multi-task training and task space pruning. In Figure 7, We compared simulation costs of DDPG baseline, multi-task DDPG with full task set, and RobustAnalog (multi-task DDPG with pruned task set). DDPG took over 300,000 simulations to pass all corner tests. With the multi-task training, the number of simulations was reduced to 35,000. With the pruned task space, the number of simulations was further cut down to 7,000. We also visualize the corner performances and the optimization trace in the performance plane of three circuit benchmarks in Figure 8.

Noise (n) - Delay (sd) plane is chosen for strongARM Latch and Bandwidth (ugb) - Phase Margin (phm) for OTAs. Selected training tasks are denoted by black circles. We can clearly see that selections are located at the performance boundary. Two snapshots of the performance distribution during the optimization are also showed. They clearly indicate that distributions moved towards the feasible set area from $t_0$ to $t_1$ with such pruned task space.

**Scale to Large Corner Sets..** Here we empirically study how the simulation cost scales as we take on more and more corner tasks. In the previous sections, we discussed the fully factorial corner test for each benchmark. In industry-level circuits, randomly sampled corners, Monte Carlo corners, are also used. There can be hundreds, even thousands of Monte Carlo corners needed to perform a thorough verification. Therefore, the scalability to a large corner set is important. To demonstrate the scalability of RobustAnalog, we conduct Monte Carlo sampling on process variation modelsets {TT, FF, SS, FS, SF}, continuous voltage range [1.0, 1.2] and continuous temperature range [0°C, 100°C] and form 5 Monte Carlo corner test sets of different sizes. These Monte Carlo corner sets have 20, 40, 80, 100, 150 corners, respectively. Experiments are done on Two-Stage OTA benchmark and results are shown in Figure 9. RobustAnalog only needs 69% more simulations when the corner task set becomes 7.5× larger.

The simulation cost difference between RobustAnalog and the baseline methods will become 4.4× larger at the scale of 150 corners.

## 5 CONCLUSION

We present RobustAnalog, a fast variation-aware optimization framework based on multi-task RL. The key property of RobustAnalog is the ability to conduct efficient multi-task learning with pruned training task space. Therefore, it can effectively design circuits for variations. We show that RobustAnalog can reduce simulation cost by an order of magnitude compared with baselines. It can also scale to a large number of variation cases. As today's chip design becomes extremely challenging with the presence of variations, RobustAnalog shows the potential to drastically shorten the circuit design cycle and reduce the cost.